# Ferri-chiral compounds with potentially switchable Dresselhaus spin splitting

Pu Huang[1,7], Zhiguo Xia[2,7,*], Xiaoqing Gao[1], James M. Rondinelli[3], Xiuwen Zhang[1,*], Han Zhang[4], Kenneth R. Poeppelmeier[5,*] and Alex Zunger[6,*]

[1]Shenzhen Key Laboratory of Flexible Memory Materials and Devices, College of Physics and Optoelectronic Engineering, Shenzhen University, Shenzhen, 518060, China

[2]State Key Laboratory of Luminescent Materials and Devices and Institute of Optical Communication Materials, South China University of Technology, Guangzhou, 510641, China

[3]Department of Materials Science and Engineering, Northwestern University, Evanston, Illinois, 60208, USA

[4]Collaborative Innovation Center for Optoelectronic Science & Technology, Key Laboratory of Optoelectronic, College of Physics and Optoelectronic Engineering, Shenzhen University, Shenzhen, 518060, China

[5]Department of Chemistry, Northwestern University, 2145 Sheridan Road, Evanston, Illinois, 60208-3113, USA

[6]Renewable and Sustainable Energy Institute, University of Colorado, Boulder, Colorado, 80309, USA

[7]These authors contributed equally to this work.

*E-mail: Alex.Zunger@colorado.edu; xiuwenzhang@szu.edu.cn; xiazg@ustb.edu.cn; krp@northwestern.edu

**Abstract**

Spin splitting of energy bands can be induced by relativistic spin-orbit interactions in materials without inversion symmetry. Whereas polar space group symmetries permit Rashba (R-1) spin splitting with helical spin textures in momentum space, which could be reversed upon switching a ferroelectric polarization *via* applied electric fields, the ordinary Dresselhaus effect (D-$1_A$) is active only in materials exhibiting nonpolar noncentrosymmetric crystal classes with atoms occupying exclusively non-polar lattice sites. Consequently, the spin-momentum locking induced by D-$1_A$ is not electric field-switchable. An alternative type of Dresselhaus symmetry, referred to as D-$1_B$, exhibits crystal class constraints similar to D-$1_A$ (all dipoles add up to zero), but unlike D-$1_A$, at least one polar site is occupied. We find that this behavior exists in a class of ferri-chiral crystals, which in principle could be reversed upon a change in chirality. Focusing on alkali metal chalcogenides, we identify $NaCu_5S_3$ in the nonentiamorphic ferri-chiral structure, which exhibits $CuS_3$ chiral units differing in the magnitude of their Cu displacements. We then synthesize $NaCu_5S_3$ (space group $P6_322$) and confirm its ferri-chiral structure with power x-ray diffraction. Our electronic structure calculations demonstrate it exhibits D-$1_B$ spin splitting as well as a ferri-chiral phase transition, revealing spin splitting interdependent on chirality. Our electronic structure calculations show that a few percent biaxial tensile strain can reduce (or nearly quench) the switching barrier separating the monodomain ferri-chiral $P6_322$ states. We discuss what type of external stimuli might switch the chirality so as to reverse the (non-helical) Dresselhaus D-$1_B$ spin texture, offering an alternative to the traditional reversal of the (helical) Rashba spin texture.

# I. INTRODUCTION

Spin splitting in nonmagnetic materials arises fundamentally from a few symmetry rules [1] in the presence of spin-orbit coupling (SOC), combining to produce an internal magnetic field that can split energy bands just as external magnetic field would do. The various forms of spin splitting in noncentrosymmetric (Non-CS) compounds [1] is generally classified into Rashba [2] (R-1) and Dresselhaus [3] (D-1) effects, whereas hidden spin splitting in centrosymmetric compounds [1] corresponds to R-2 and D-2, where R and D are distinguished from each other by their helical and non-helical spin textures, respectively. The allure of such effects lies in the hope that the relevant spin splitting can be switched by an external operation, to the benefit of spintronic devices. This generally requires some spatial directionality of the underlying effect—such as vector dipolar polarization—that can be manipulated in a device structure. Such directionality is supplied by the R-1 Rashba effect, but not by the conventional D-1 Dresselhaus effect, since atoms occupy nonpolar sites in D-1 crystals. Indeed, the R-1 effect is enabled in Non-CS compounds that have a polar point group due to a nonzero sum of local dipoles from atoms occupying polar sites. Thus, the R-1 effect is associated with a net directional electric dipolar polarization [1-2, 4]; once the ferroelectric polarization is switched, e.g., by an electric field, the Rashba spin splitting can be reversed [5]. On the other hand, the traditional Dresselhaus effect [1, 3-4] (hereby denoted D-$1_A$) is enabled in Non-CS compounds where each occupied Wyckoff orbit has nonpolar site symmetry, and dipoles add up to zero, as in tetrahedral GaAs. As this D-$1_A$ effect is created by a non-directional inversion asymmetry, external electric fields cannot reverse the spin texture.

Zhang et al. noted [1] that the definition of different bulk spin splitting functionalities determined by the point group and space group symmetries depends on additional aspects of the site symmetries of the individual Wyckoff positions occupied by the atoms in the structure. Although the Dresselhaus effect requires a nonpolar Non-CS crystal class for the space group [6], it can be further differentiated into 3 sub-classes based on the occupied site symmetries in the crystal [1]. In the traditional D-$1_A$ effect, each occupied site must be nonpolar, whereas in D-$1_B$ at least one site must be polar, however, the net electric polarization must be zero, i.e., the sum of dipoles arising from local displacements must be zero. In contrast, D-$1_C$ polar sites must be occupied, but the sum of dipoles is nonzero. Herein we focus on the D-$1_B$ effect, which although similar to that in the Rashba effect differs from the latter in that there is zero net electric polarization and the spin texture is non-helical. Chiral crystals with polar individual site symmetries (e.g., LiAlO$_2$) are examples of D-1$_B$ materials, which compensate each other even though they are not inversion partners. This special Dresselhaus spin splitting enabled by crystallographic chirality is rarely studied, but could also be used to remove the Kramer's degeneracy in crystals with relativistic interactions and access spin-polarized carriers in nonmagnetic compounds.

Here, we focus on ferri-chiral materials, where two sets of chiral units with opposite distortion senses cannot exactly compensate each other (Fig. 1). The direction of these small atomic displacements governs the spin-split electronic band structure, which can be reversed upon changing the orientations of these units. We identify NaCu$_5$S$_3$ as a stable ferri-chiral compound by symmetry screening in the diverse family of alkali metal copper chalcogenides followed by analysis of the spin-splitting in the electronic structure with density functional calculations. We then synthesize NaCu$_5$S$_3$, and computationally demonstrate that the Dresselhaus spin splitting is interlocked with its ferri-chirality. We show that application of modest strain significantly reduces the barrier between the two chiral phases. We discuss features of various external stimuli that might switch the chirality and reverse the (non-helical) Dresselhaus D-1$_B$ spin texture, offering an alternative to the traditional reversal of the (helical) Rashba spin texture.

## II. THEORETICAL SEARCH FOR FERRI CHIRAL COMPOUNDS OF TYPE D-1$_B$

We search for compounds that satisfy the basic design principles for ferri-chiral D-1$_B$ Dresselhaus spin splitting. These design constraints include: (i) the material is chiral with nonpolar and noncentrosymmetric space groups; (ii) at least one atomic site in the unit cell of the crystal has polar site symmetry; (iii) the transition barrier between the two chiral phases is sufficiently low, (e.g. does not require bond breaking); and (iv) the material is thermodynamically stable (i.e. does not phase separate).

Regarding the last condition, we previously advanced an approach to search for materials that are stable and likely synthesizable by constructing the chemical potential polygon [7-9], in the relevant elemental chemical potential space. This process relies on finding whether there is a region ("green zone") where none of the possible competing phases will render the targeted phase unstable. Details of the density functional theory used are given in Appendix I. Application to the equiatomic (p=q=r=1) ternary A$_p$B$_q$X$_r$ 18–electron phases has previously shown numerous [7, 10], never before made compounds that were subjected to the search of target functionalities and selective synthesis [Fig. 2(a-c)]. Here, we extend the search of stable new phases to general, i.e., non-equiatomic stoichiometries (p, q, r) of the ternary {A, B, X} family with $A^I$, $B^I$ and $X^{VI}$, filtering compounds that fulfill the conditions for ferri-chiral D-$1_B$ Dresselhaus spin splitting. To find the lowest-energy crystal structure of a missing compound (e.g., NaCuS), we construct a set of candidate crystal structures for its composition from those that are known for existing compounds in ICSD [33], such as AgMgAs-type ($F\bar{4}3m$), LiGaGe-type ($P6_3mc$) and ZrBeSi-type ($P6_3/mmc$) structures for 1:1:1 composition, and compute the total energy of each missing material in each structure subject to local relaxations, and then select from this list the lowest-energy structure [34]. For the reported compounds, we use the experimentally determined crystal structures. To determine the thermodynamic stability of a target ternary compound $A_lB_mX_n$, we consider all of its possible disproportionation channels to the competing phases (elemental phases, binaries and ternaries) with neighboring compositions. This is done by solving a set of inequalities in a 3-dimensional space of chemical potentials $\Delta\mu_A$, $\Delta\mu_B$ and $\Delta\mu_X$. For all competing elemental phases, we used inequalities of the form $\Delta\mu_I$ (I = A, B, X) < 0. For all competing binary phases $A_pB_q$, we used the inequalities $p\Delta\mu_A + q\Delta\mu_B < \Delta H_f (A_pB_q)$. For all competing ternary phases $A_pB_qX_r$, we used $p\Delta\mu_A + q\Delta\mu_B + r\Delta\mu_X < \Delta H_f (A_pB_qX_r)$. This 3-dimensional space of chemical potentials was reduced to 2 dimensions by the $l\Delta\mu_A + m\Delta\mu_B + n\Delta\mu_X = \Delta H_f (A_lB_mX_n)$ equation. The values of $\Delta\mu_A$ and $\Delta\mu_B$ where all inequalities are satisfied define the stability area of the $A_lB_mX_n$ compound.

Our analysis of sodium thiocuprates in the Na-Cu-S system shows several previously made I-I-VI compounds indicated by check marks in Fig. 2(a) and previously unknown compounds predicted here are stable (denoted by + sign; subsequently realized in experiments [17-18] except NaCuS). Appendix II lists the predicted as well as previously known alkali metal copper chalcogenides. According to structural symmetry analysis, we identify one compound ($NaCu_3Te_2$) with a polar space group permitting R-1 spin splitting, three materials (NaCuS, KCuSe, and KCuTe) with hidden spin polarization (D-2) effect [1], and 26 compounds (e.g., $KCu_3Te_2$) with the R-2 and D-2 effects [1]. We identify among alkali metal copper chalcogenides, one candidate compound $NaCu_5S_3$ that possesses the structural symmetries required for the D-$1_B$ effect.

We evaluated the thermodynamic stability of the candidate material $NaCu_5S_3$ against disproportionation into its competing phases, as well as those of the other ternaries in Na-Cu-S system, to examine the synthetic feasibility. Our results show that $NaCu_5S_3$ exhibits the largest stability (green) zone [0.287 eV$^2$, Fig. 2(c)]: it is a few times larger than those of NaCuS [0.08 eV$^2$, Fig. 2(b)] and the other ternaries. Our first-principles thermodynamic data agree well with the previous synthesis [11-16] of Na-Cu-S compounds, and the calculated sizes of stability zones suggest that the synthetic feasibility of NaCu$_5$S$_3$ is the highest among all Na-Cu-S compounds.

## III. SYNTHESIS AND CHRYSTALLOGRAPHIC ANALYSIS

### A. Synthesis

Six sodium thiocuprates in the Na-Cu-S system, including NaCu$_5$S$_3$, NaCu$_4$S$_4$, Na$_2$Cu$_4$S$_3$, Na$_3$Cu$_4$S$_4$, Na$_4$Cu$_2$S$_3$ and Na$_7$Cu$_{12}$S$_{10}$ [11-16], were generally prepared by melting mixtures of the following three products in different ratios of sodium metal (sodium sulfide, Na$_2$S), copper (copper sulfide, Cu$_2$S), and sulfur powder. Reagent chemicals were used as received: NaOH flake (98%, Alfa Aesar), Cu$_2$S (99.5%, Alfa Aesar), Thiourea (99%+, Sigma-Aldrich), Ethanediamine ($\geq$ 99%, Sigma-Aldrich).

We prepared NaCu$_5$S$_3$ microcrystals by using the solvothermal method and ethanediamine as the solvent. NaCu$_5$S$_3$ was first reported by Effenberger and Pertlik [11] in 1985 *via* the hydrothermal synthesis from the starting materials of Cu powder and Na$_2$S in H$_2$O. Here, we prepared the NaCu$_5$S$_3$ microcrystals by a facile solvothermal method. In our typical procedure, 3.125 mmol Cu$_2$S, 0.625 mmol Thiourea, 12.5 mmol NaOH (10 fold excessive stoichiometry) and 6 mL ethanediamine were added in a 3 × 1.75 inches rectangle Teflon pouch. Then six Teflon pouches can be placed in a 100 mL Teflon-lined Parr pressure vessel filled with 50 mL of deionized H$_2$O as backfill. Pressure vessels were heated to 200 °C for 24 h and cooled to room temperature naturally. Pouches were opened in air, and the microcrystal products were recovered *via* vacuum filtration and the following drying in a vacuum furnace at 60 °C for 12 h.

### B. Characterization

Phase purity of the products was assessed by powder XRD (PXRD). The products were finely ground and mounted on a flat plate sample holder. Diffraction data were collected on a Rigaku MiniFlex 600 diffractometer with the Cu K$\alpha$ source, operating at 40 kV and 15 mA under a continuous scanning method in the 2$\theta$ range 5−120°. PXRD data were analyzed with the Rietveld method using TOPAS 4.2 program. The morphologies of the typical as-prepared NaCu$_5$S$_3$ microcrystal product and the cracked section of the NaCu$_5$S$_3$ pellet were performed with a Hitachi S-4700-II scanning electron microscope.

Electrical conductivity measurements were performed on as-prepared NaCu$_5$S$_3$ microcrystal products were firstly densified by SPS method (SPS-211LX, Fuji Electronic Industrial Co., Ltd.), which was treated at 723 K for 5 min in a 12.7-mm-diameter graphite die under an axial compressive stress of 40 MPa in vacuum. Highly dense (>95% of theoretical density) disk-shaped pellets with dimensions of 12.7 mm diameter and 8 mm thickness were obtained.

## C. Crystallographic Analyses

Fig. 3(a) presents the results of our powder x-ray diffraction measurements. All diffraction peaks were indexed using a hexagonal cell ($P6_322$) with parameters close to previously reported NaCu$_5$S$_3$, and the refined lattice constants, a = 7.0102 (1) Å, c = 7.2395 (1) Å and V = 308.09 (1) Å$^3$. Our Rietveld refinements yielded excellent refinement statistics, $R_{wp}$ = 3.03%, $R_p$ = 1.88%, $R_{exp}$ = 1.03%, and $\chi^2$ = 2.96, which indicate that the sample has been prepared with high phase purity. Well-defined NaCu$_5$S$_3$ crystals varying in size from 30 to 50 μm with rod- and plate-like morphologies are observed [see the SEM image in Fig. 3(b)], which are consistent with the crystalline habit of the hexagonal crystal system. Exposed to SPS sintering at 500 °C, the NaCu$_5$S$_3$ microcrystals are fractured into smaller uniform ones with 2-3 μm diameters in size [Fig. 3(c)]. We observe compact packing after the SPS process, which supports the high relative-density 96% calculated by volume and weight.

NaCu$_5$S$_3$ belongs to the class of chiral materials with a nonpolar nonentiamorphic space group ($P6_322$, No.182), meaning that the same space group can permit structures in two equivalent descriptions rather than forming enantiomorphic pairs (e.g., $P6_122$ and $P6_522$). The crystal structure for the experimentally synthesized phase is shown in Fig. 3(d) and the occupied Wyckoff sites are presented in Appendix II Table A1. The important feature emphasized herein, following Zhang et al. [1], is that the point group polarities of individual Wyckoff positions matter: the Na sub-lattice in NaCu$_5$S$_3$ has the Non-CS and nonpolar point group $D_3$; The two Cu sub-lattices host $C_3$ (polar and chiral) and $C_2$ (polar and chiral) site symmetries, respectively, and the sum of dipoles on the Cu sub-lattices is zero. Although the Cu sub-lattices are individually polar, they cannot contribute to a net polarization because the axis of rotation for the site symmetry is orthogonal to the direction of the displacements permitted by the free parameter in the Wyckoff orbits. This is the key local crystallographic feature that enables the D-1$_B$ Dresselhaus subclass without unlocking a Rashba-like interaction. Finally, the S sub-lattice has the polar chiral site symmetry $C_2$, which also prohibits a net dipole.

We next use these refined experimental structures with additional symmetry analysis to further understand the atomistic origin of the chirality in NaCu$_5$S$_3$, which we show is best referred to as a ferri-chiral configuration. NaCu$_5$S$_3$ can exist in two forms: the higher-energy symmetric CS achiral $P6_3/mcm$ structure [Fig. 4(a, b)], which can transform into one of the two lower-energy chiral configurations FEC-1 and FEC-2 of space group $P6_322$ [Fig. 4(c, d)] possessing a Non-CS space group and individual polar sites with dipoles that sum to zero. Our synthesis obtained the FEC-1 configuration as in previous work [11], leaving the other chiral configuration (FEC-2) yet to be realized. The chiral motifs and the mechanism of ferri-chirality of NaCu$_5$S$_3$ is described as follows:

(a) The chiral copper sulfide structural motifs are linearly arranged along $c$ axis in NaCu$_5$S$_3$ crystal [Fig. 4(b)], with alternating left-handed chirality ($\sigma^+$) and right-handed chirality ($\sigma^-$). The smallest chiral unit consists of two CuS$_3$ triangles arranged one above the other with a relative rotation of 60 degree, and the CuS$_2$ chains connecting their vertices. This chiral unit can be deduced from a triangular prism by a right-handed (left-handed) 'twist', hinting at its chirality.

(b) The Non-CS chiral phase of NaCu$_5$S$_3$ possesses $P6_322$ symmetry and enantiomorphic point group 622 ($D_6$). Owing to the sense of the net atomic displacements within the lattice, two chiral configurations (domains) with $P6_322$ symmetry are possible in the same space group [Fig. 4(c)]: the left (right) structure exhibits the $\sigma^-$ ($\sigma^+$) chiral structural units shrinking and the $\sigma^+$ ($\sigma^-$) chiral structural units expanding, leading to a net ferri-chirality with partially compensated opposite handednesses.

(c) The CS achiral phase of NaCu$_5$S$_3$ as shown in Fig. 4(a) belongs to the $P6_3/mcm$ space group (see Table A1 for its Wyckoff positions). Since the CuS$_3$ triangles in the chiral units become ideally planar and the CuS$_2$ chains evolve into ideally linear geometries, the $\sigma^+$ and $\sigma^-$ units exactly compensate, leading to an achiral CS phase.

We now evaluate the existence of the ferri-chiral state as found in experiment and theory, and determine the atomistic displacements that enable the ferri-chirality. This process involves displacements of Cu-1 in CuS$_2$ chain, [Fig. 3(d)] and Cu-2 (in CuS$_3$ triangle) that can transform as the irreducible representation $\Gamma_1^-$. Although CuS$_2$ linear chains have been found in copper chalcogenides [33], CuS$_3$ planar triangles are an uncommon coordination for copper, which provides a driving force for bond distortion to a non-planar structure as follows: Cu-1 displaces within the *ab*-plane, which removes one mirror plane, whereas Cu-2 displaces along the *c* axis, eliminating the reflection operation perpendicular to the 6-fold axis. Chirality arises in this structure because of a perceptual rotation sense due to the anti-symmetric displacements on the copper planes as illustrated for the ferri-chiral phase of NaCu$_5$S$_3$ with red/blue arrows in Fig. 4(c), which together remove the glide operation. Indeed, these local 3-in-2-out (or 3-out-2-in) displacements exhibit dihedral symmetry; however, their supposition results in partial instead of full compensation of the $\sigma^+$ and $\sigma^-$ chiralities to produce the ferri-chiral structures with $P6_322$ symmetry.

## IV. BAND DISPERSIONS AND SPIN TEXTURES INCLUDING SOC

We next evaluate the spin splitting and spin texture of the ferri-chiral phases of NaCu$_5$S$_3$ by DFT+$U$+SOC with $U$ = 6 eV (Fig. 5). The Non-CS $P6_322$ ferri-chiral phase show strong spin splitting at the top valence bands along the $\Gamma-A$ direction, in contrast to the CS $P6_3/mcm$ achiral phase with no spin splitting. Band structures and electronic densities-of-states obtained without SOC are shown in Appendix III Fig. A1 and Fig. A2, respectively. Fig. A3 compares the band dispersions without SOC obtained using a screened hybrid functional (HSE06 [41, 42]) against our DFT+$U$ results, showing good agreement between the two levels of theory. Fig. A4 shows the effect of different choices of $U$ on the band structure without SOC.

Detailed band structures near the valence band maximum (VBM) in Fig. 5(c−e) show the lifting of the Kramer's degeneracy and the appearance of spin splitting in the ferri-chiral phases, as compared to the CS achiral phase. The overall orbital energy levels dependencies found in Fig. 5(a) are similar to the electronic structure without SOC [Fig. A1], because of the broken inversion; however, the Kramer's degeneracies are only fully removed in the former case with SOC, which results in spin splitting that is clearly discernable at the VBM along the $\Gamma-A$ direction in Fig. 5(a) [see Fig. 5(c) for the spin notation]. Fig. 6 and Appendix II Fig. A2 shows that the following atomic orbitals are hybridized and mainly contribute to the valence bands along $\Gamma-A$ direction: S $p_x$, $p_y$, Cu-2 $d_{x^2-y^2}$, $d_{xy}$, Cu-1 $d_{yz}$, $d_{xz}$ and $d_{z^2}$.

Although the two ferri-chiral structures possess the same crystalline symmetry, the spin splitting and spin textures are significantly distinct. For FEC-1 phase of NaCu$_5$S$_3$, the top two valence bands along the Γ−A *k*-path split into four singly degenerate bands with spin polarization along the *z* direction and individual spin components, showing large linear Dresselhaus spin splitting [Fig. 5(c)] with linear spin splitting coefficient of ($\frac{1}{2}\frac{dE}{dk} \cong 275$ meV·Å). The derived spin splitting energy for the top valence bands near the A point is ~208 meV, which is remarkably large for the low atomic mass elements in NaCu$_5$S$_3$. Whereas in the FEC-2 phase, the spin splitting is reversed as compared to FEC-1, producing completely opposite spin components [Fig. 5(e)]. In other words, the spin splitting switches between the two ferri-chiral phases of NaCu$_5$S$_3$, suggesting that the ferri-chirality is associated with the Dresselhaus spin splitting (D-1$_B$ type), analogous to the association between dipolar polarization and R-1 spin splitting. Similarly, Fig. 7 shows the spin split bands in the $k_y$-$k_z$ plane ($k_x$ = 0) with annotated spin polarization directions for the FEC-1 phase, demonstrating the Dresselhaus-like non-helical spin textures. (The corresponding spin texture of FEC-2 phase is opposite to that of FEC-1).

## V. PROSPECTS FOR SWITCHABILITY OF FERRI CHIRAL SPIN TEXTURE

The ferri-chiral phases in NaCu$_5$S$_3$ differ only by chiral unit contracting or expanding without breaking of bonds [Fig. 4(c, d)], thus the transition barrier between such phases could be rather low. We assess computationally the transition barrier from FEC-1 to FEC-2. The conversion between the two ferri-chiral phases [indicated by the red/blue arrows in Fig. 4(c)] involves only small atomic displacements, but not breaking of chemical bonds or twisting of structural units as in a reconstructive transformation, suggesting a low energy barrier for the monodomain state transformation. The calculated transition barrier between FEC-1 and FEC-2 is as low as ~12 meV/f.u. as indicated by the black dashed curve (unstrained case) in Fig. 8a.

***Yet, unlike the case in R-1 compounds, here a static and conventional electric field alone cannot switch the polarization. Another switching agent needs to be identified.*** At this time we note that switching of chirality in molecular materials has been realized in experiments using external stimuli such as photo-excitation [19-21], electron tunneling [22], and electric field [23]. On the other hand, chirality-selective synthesis, i.e. chiral induction, was applied to both molecular materials [24-26] and inorganic crystals [27], by using circular polarized light. Furthermore, circularly polarized X-ray [28] and light [29-31], as well as optical vortex [32], were widely used to probe/manipulate materials' chirality. Whether these approaches can be used or new avenues advanced for switching NaCu$_5$S$_3$ is a challenge for future research.

The ferri-chiral transition barrier can be tuned to zero by a few percent biaxial tensile strain, or increased significantly by compressive strain [Fig. 8(a)]. indicating thin film, device relevant geometries of NaCu$_5$S$_3$, probed with electrical measurements or optical spectroscopies are a promising strategy to demonstrate the ferri-chiral dependent spin splitting but not necessarily switch it. Appendix II Table A2 lists the potential substrate material for imposing tensile strain on NaCu$_5$S$_3$, which may be grown using pulse-laser deposition from ceramic targets our synthesis strategy for high phase purity. For large enough biaxial tensile strain (~3%), the atomic displacement that breaks the full compensation of $\sigma^+$ and $\sigma^-$ chiralities can be restored [see the variation of the crystallographic parameters in Fig. 8(b)], and the CS anti-chiral phase ($P6_3/mcm$) is stabilized as the ferri-chiral transition barrier is quenched.

The ferri-chirality and D-1$_B$ Dresselhaus spin splitting in NaCu$_5$S$_3$ are inherently interlocked. Fig. 5(c, e) demonstrates the switch of spin splitting once the chirality is reversed. Furthermore, as the D-1$_B$ effect is due to a convergent effect of the atomic site dipolar polarizations (that sum to zero in the ferri-chiral structure) and thus depends on their relative positions and magnitudes. For that reason, the D-1$_B$ spin splitting could be much more sensitive to crystallographic parameters than the D-1$_A$ effect—this sensitivity could offer new knobs to manipulate the electron's spin. Fig. 8(c, d) show that the spin splitting energy near the A point of NaCu$_5$S$_3$ can be tuned from 16 meV under 3% tensile strain along the $c$ lattice direction to 435 meV under 3% compressive strain. As to the biaxial strain in the $ab$ plane, along with the significant modulation of the chirality-enabling atomic displacements [Fig. 8(b)], a wide range of spin splitting energy from complete zero spin splitting to ~200% of the unstrained case near A point can be obtained (see Fig. A5 for the electronic structures of NaCu$_5$S$_3$ under biaxial/uniaxial strains along $a/b$ lattice directions). The calculated strong strain modulation effect of D-1$_B$ spin splitting further demonstrates the interlocking of ferri-chirality and Dresselhaus effect in ferri-chiral materials, which could offer chirality-related modalities to manipulate/detect spin splitting, or *vice versa*.

## VI. CONCLUSION

We predict a potentially switchable Dresselhaus spin splitting in the proposed ferri-chiral structural symmetry exhibited by NaCu$_5$S$_3$. The non-conventional Dresselhaus effect D-1$_B$ found in the experimentally synthesized candidate compound NaCu$_5$S$_3$ is interlocked with its ferri-chirality, analogous to the association of the Rashba effect with dipolar polarization. The ferri-chiral transition barrier can be significantly lowered or quenched by external strain, facilitating the potential transition between monodomain ferri-chiral states as well as their associated spin splitting by external stimuli. Yet, unlike the case in R-1 compounds, here electric field cannot switch the polarization. At this time we note that switching of chirality in molecular materials has been realized in experiments using external stimuli such as photo-excitation [19-21], electron tunneling [22], and electric field [23]. On the other hand, chirality-selective synthesis, i.e. chiral induction, was applied to both molecular materials [24-26] and inorganic crystals [27], by using circular polarized light. Furthermore, circularly polarized X-ray [28] and light [29-31], as well as optical vortex [32], were widely used to probe/manipulate materials' chirality. Exploring such avenues for switching $NaCu_5S_3$ is a challenge for future research. This study initiates the design of novel ferri-chiral materials with potential applications in spintronics and optoelectronics: the Non-CS ferri-chiral phases are active for piezoelectric, nonlinear optics, as well as circularly polarized light that interacts with spin splitting, offering a natural platform to combine mechanical, optical, and spintronic effects. Furthermore, the association of Dresselhaus spin splitting with a structural chirality enables a route to deliver large Dresselhaus spin splitting beyond the strength of SOC. Last, the reversibility of non-helical Dresselhaus spin splitting, being more complicated than the helical Rashba spin texture, could offer plural spin information from a single spintronic device operation.

**Acknowledgements**

CU and NU was supported by NSF-DMR EPM program under grant DMR-1806939 and DMR-1806912, respectively. J.M.R. acknowledges funding from the Army Research Office (W911NF-15-1-0017). The work in China was supported by National Natural Science Foundation of China (Grant No. 11774239, 11804230, 51722202, 61827815), National Key R&D Program of China (Grant No. 2016YFB0700700), and Shenzhen Science and Technology Innovation Commission (JCYJ20170818093035338, JCYJ20170412110137562, ZDSYS201707271554071).

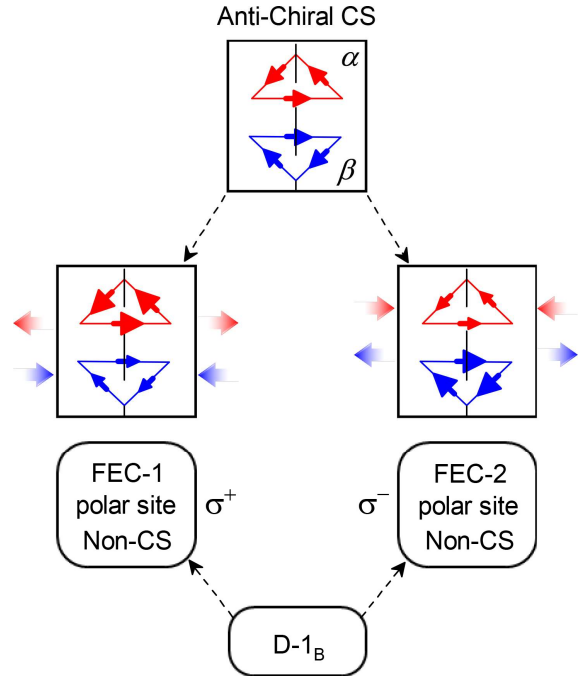

**FIG. 1.** Schematic depiction of how achiral centrosymmetric (CS) symmetry can decompose into two ferri-chiral (FEC) structures FEC-1 and FEC-2 that are Non-CS. The special case discussed here occurs when the FEC structures exhibit the Dresselhaus effect D-$1_B$ where at least one site has polar symmetry. This behavior is distinct from the standard Dresselhaus effect D-$1_A$ where all sites are non-polar, as in the zinc blende structure.

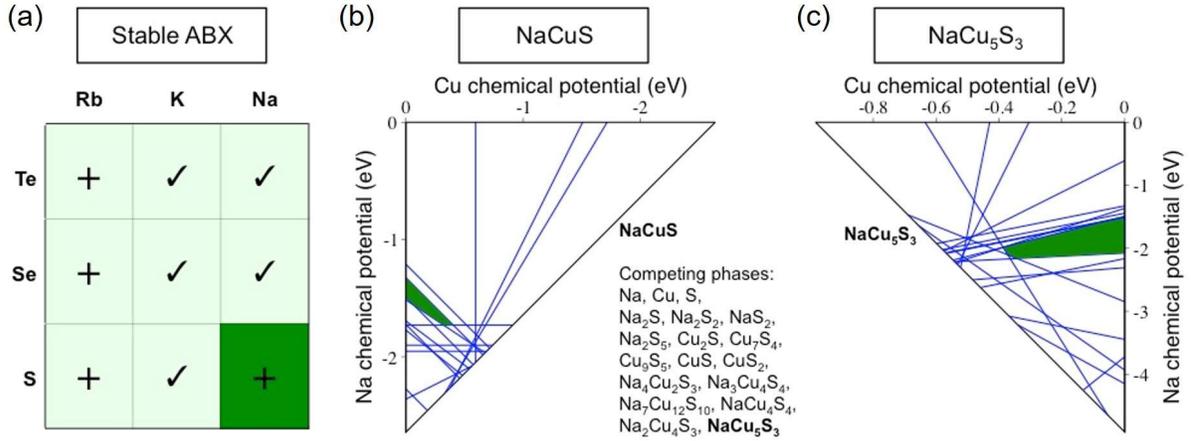

**FIG. 2.** Stability assessment of alkali metal copper chalcogenides. (a) Predicted (+ signs) and previously reported (check marks) I-I-VI compounds. (b, c) Stability regions (green zones) of NaCuS and NaCu$_5$S$_3$ in the space of elemental chemical potentials. Each blue line indicates a binary or ternary competing phase. The areal size of the stability regions (in units of eV$^2$) for NaCuS and NaCu$_5$S$_3$ are 0.080 and 0.287, respectively, compared to the other ternary compounds in the Na-Cu-S system: Na$_4$Cu$_2$S$_3$ (0.175), Na$_3$Cu$_4$S$_4$ (0.083), Na$_7$Cu$_{12}$S$_{10}$ (0.052), NaCu$_4$S$_4$ (0.037), and Na$_2$Cu$_4$S$_3$ (0.003).

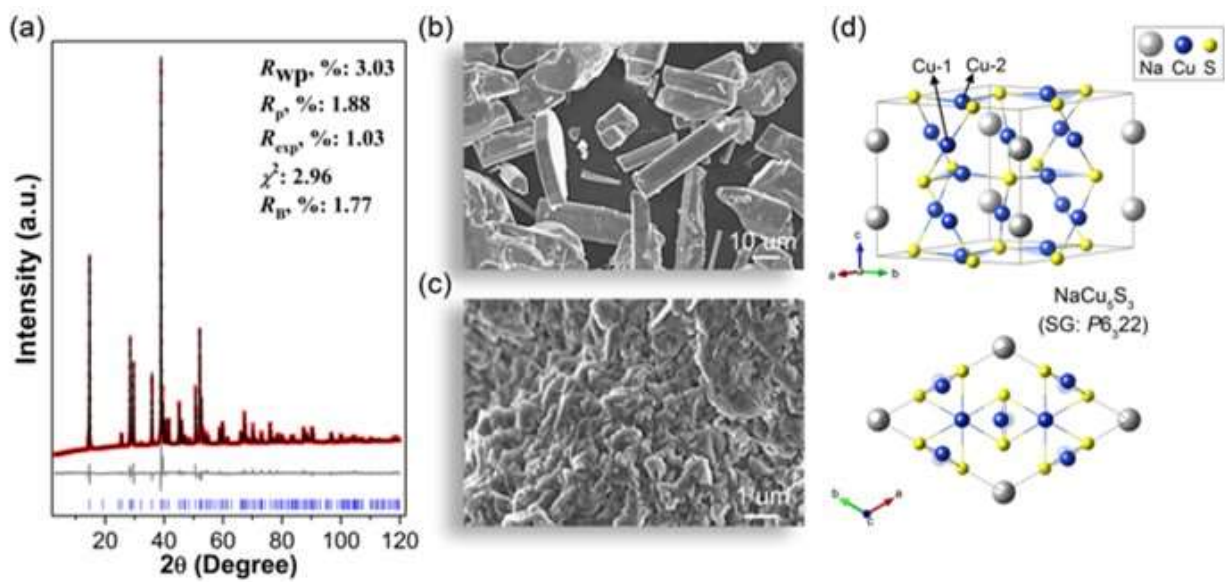

**FIG. 3.** Experiment synthesized NaCu$_5$S$_3$. (a) Rietveld fit to PXRD data. (b) Typical SEM image of as-prepared NaCu$_5$S$_3$ microcrystals. (c) The section SEM image of the bulk NaCu$_5$S$_3$ obtained by the SPS sintering at 500 °C with high relative-density of 96%. (d) Crystal structure of NaCu$_5$S$_3$ ($P6_322$) with two unique Cu atoms (Cu-1 and Cu-2).

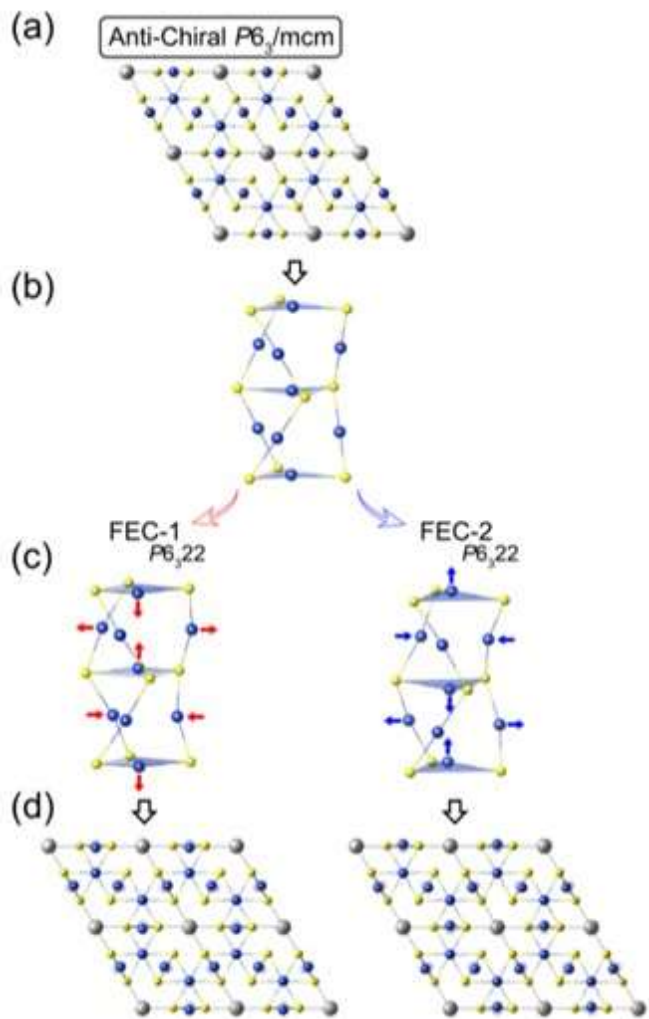

**FIG. 4.** Schematic representation of the symmetry evolution for NaCu$_5$S$_3$. (a, b) Anti-chiral (AC, fully compensated opposite chiralities). (c, d) Ferri-chiral (FEC, partially compensated opposite chiralities) phases. The "twisted triangular prism" consists of 2 CuS$_3$ and 3 CuS$_2$ motifs constitute the structural chiral unit. FEC-1 is the experimental phase subject to DFT struture relaxation. FEC-2 is deduced from FEC-1 by an inversion transformation.

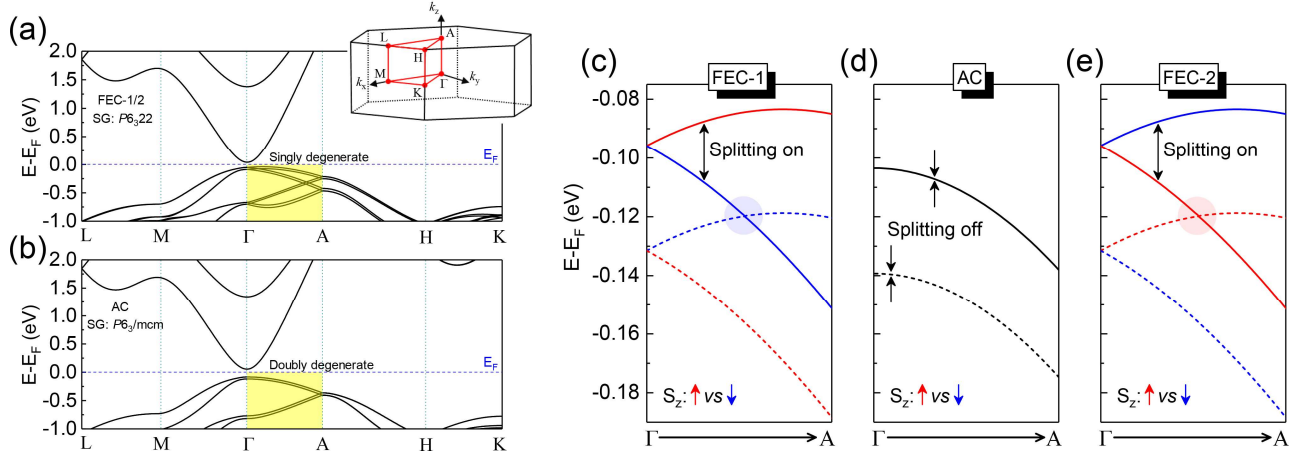

**FIG. 5.** D-1$_B$ Dresselhaus spin splitting in NaCu$_5$S$_3$. (a, b) Band structures for NaCu$_5$S$_3$ with $P6_322$ and $P6_3/mcm$ symmetries calculated by DFT+$U$+SOC. Inset: The Brillouin zone of the hexagonal structures. (c-e) Detailed band structure information near VBM along Γ−A $k$-line for the anti-chiral (AC) and ferri-chiral (FEC-1 and FEC-2) phases of NaCu$_5$S$_3$. Red/blue colors in **c** and **e** indicate the spin projection with the spin polarization axis along the $z$ direction.

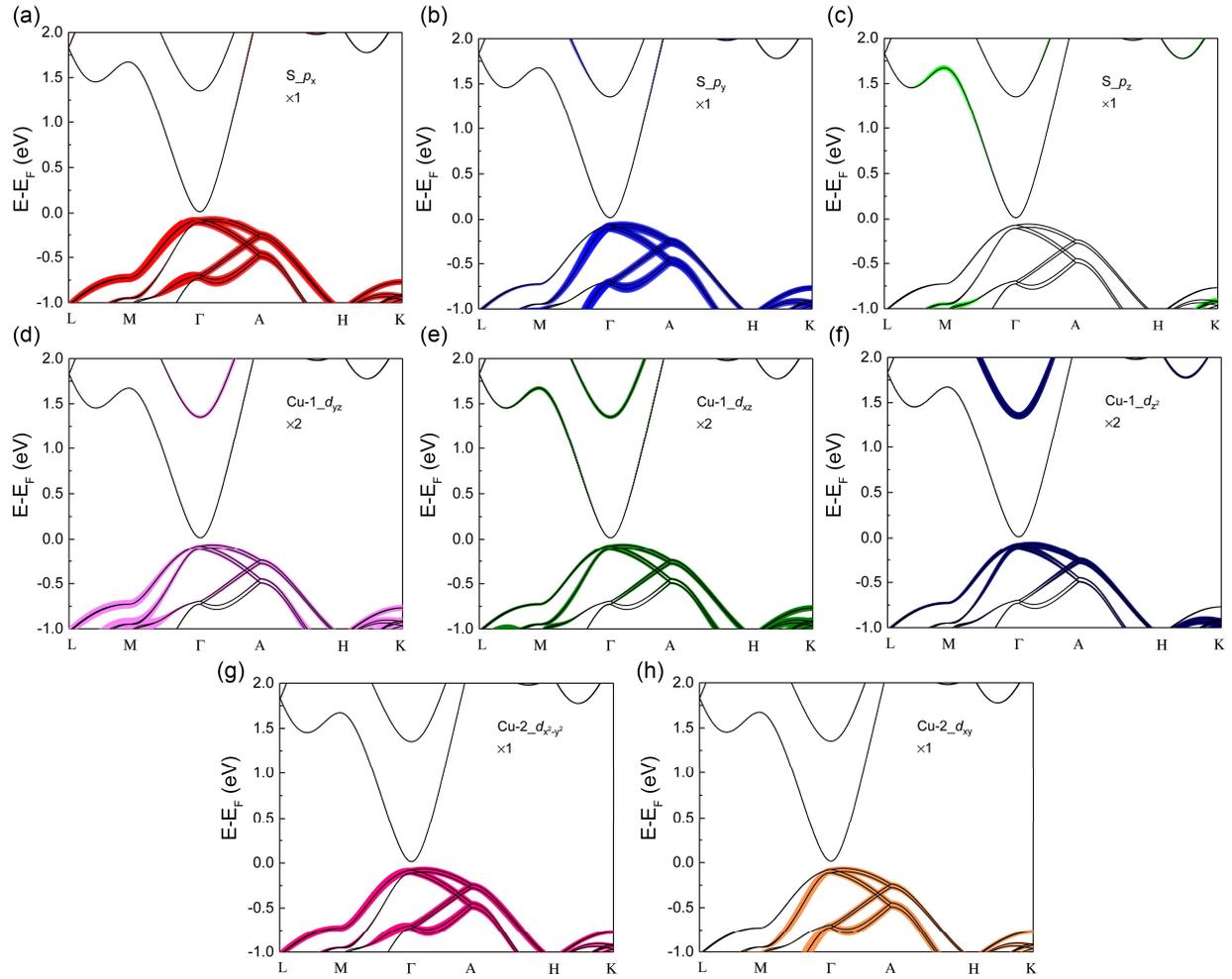

**FIG. 6.** Atom and orbital projected band structures (a-h) of NaCu$_5$S$_3$ with $P6_322$ symmetry calculated at the DFT+$U$+SOC level.

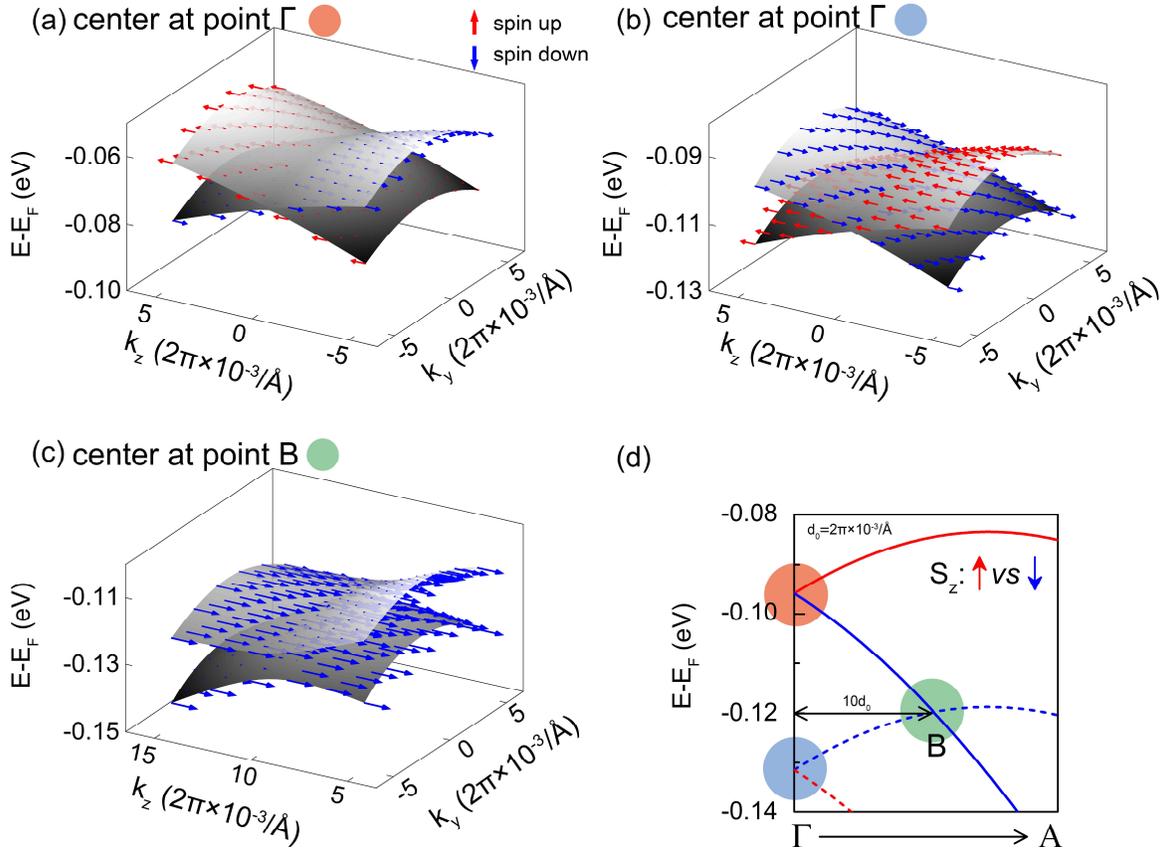

**FIG. 7.** Spin textures on the ($k_y$, $k_z$) plane ($k_x = 0$) for the top valence bands of NaCu$_5$S$_3$ ($P6_322$) at the DFT+$U$+SOC level. (a-c) Spin textures for different ranges of valence bands. (d) Zoom-in band structure showing the three chosen ranges of valence bands for spin texture illustration.

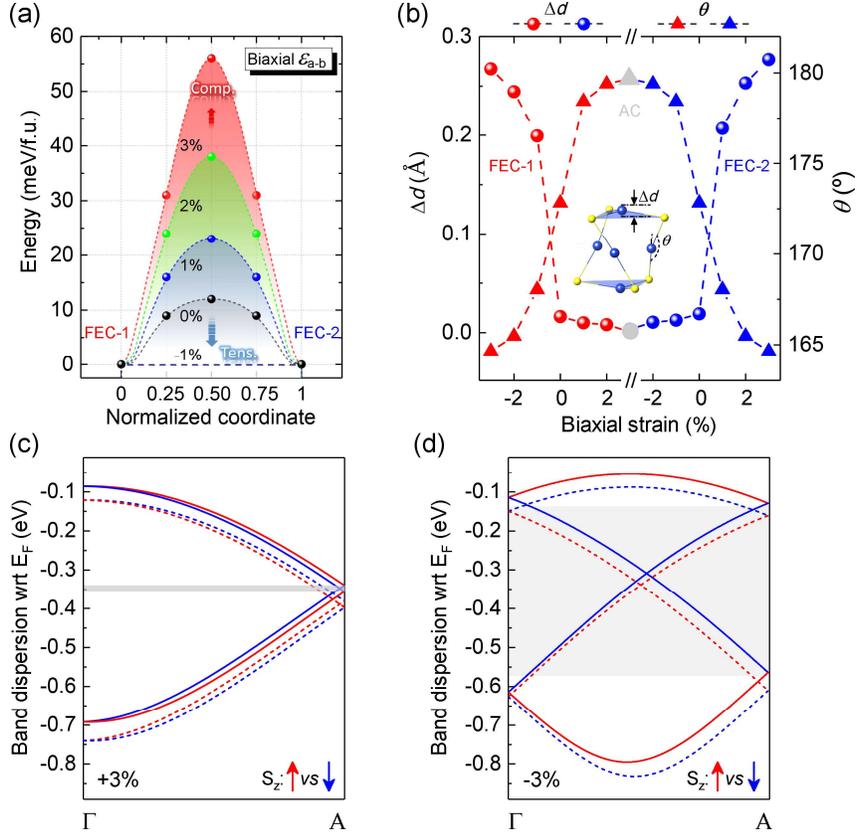

**FIG. 8.** Tunable ferri-chiral phase transition and spin splitting in $NaCu_5S_3$. (a) Transition barrier between the two ferri-chiral phases of $NaCu_5S_3$. The black dashed curve corresponds to the unstrained case with the other curves correspond to different strain values. (b) Detailed crystallographic information including the movement of Cu-2 away from the center of $CuS_3$ planar triangle along [001] lattice direction ($\Delta d$), the movement of Cu-1 away from the center of $CuS_2$ linear chain as represented by the S-Cu-S angle ($\theta$), along the phase transition between FEC-1/2 and AC states under external biaxial strain in the $ab$ plane (here the +/- signs represents tensile and compressive strain). (c, d) Low energy valence bands in the $\Gamma-A$ direction for FEC-1 ferri-chiral phase of $NaCu_5S_3$ under uniaxial strain along the $c$ lattice direction. Red/blue colors indicate the spin projection with the spin polarization axis along the $z$ direction. The gray background indicates the magnitude of the spin splitting energy. Similar tunability of spin splitting energy under $a/b/c$ uniaxial/biaxial strains is demonstrated in Appendix Fig. A5.

**Appendix I. Details of density functional calculations**

The electronic structure of NaCu$_5$S$_3$ is evaluated by density functional theory (DFT) [35] using the projector-augmented wave (PAW) pseudopotentials [36] with the exchange-correlation of Perdew-Burke-Ernzerhof (PBE) form [37] as implemented in the Vienna Ab-initio Simulation Package (VASP [36, 38]). The electronic wavefunctions were expanded using a planewave basis set with an energy cutoff of 520 eV. The Brillouin zone was sampled on 8×8×8 Monkhorst-Pack [39] $k$-point meshes. For Cu element, we introduced the on-site Coulomb interactions in the localized $d$ orbitals by using the DFT/GGA+$U$ method [40]. The initial crystal structure is taken from experiment, and the lattice constants and atomic positions are fully relaxed under the energy and force tolerance of $10^{-6}$ eV and $5\times10^{-3}$ eVÅ$^{-1}$ per unit cell using a conjugate-gradient (CG) algorithm. Spin-orbit coupling (SOC) effects are taken into account by a perturbation $\sum_{i,l,m} V_l^{SO} \boldsymbol{L} \cdot \boldsymbol{S} |l,m>_{ii}<l,m|$ to the pseudopotential, where $|l,m>_i$ is the angular momentum eigenstate of the $i$th atomic site [41]. The Wigner-Seitz radii for constructing $|l,m>_i$ used in this study are listed in the pseudopotentials of the VASP simulation package [38]. The spin state of the electron's wavefunction is evaluated by projecting the calculated wavefunction $|\phi>$ on the spin and orbital basis of each atomic site $C_{i,l,m,\eta} = <\phi|(s_\eta \otimes |l,m>_{ii}<l,m|)|\phi>$. The ferri-chiral transformation barriers are evaluated by using the nudged elastic band (NEB) method [42], during which the crystal symmetry and band structure evolution of every image were carefully depicted. We are aware of the DFT band gap error and compare the band gap obtained from hybrid functional (HSE06 [43-44]) method with that from DFT+$U$, and found that DFT+$U$ with $U$ = 6 eV on Cu predicts almost the same band gap value as HSE06. Crystal structures and charge densities were drawn using the VESTA software [45].

**Appendix II. Predicted and previously known achiral alkali metal copper chalcogenides**

The experimentally reported [11, 33] achiral alkali metal copper chalcogenide compounds in A-B-X systems (A = Na, K, Rb, B = Cu, X = S, Se, Te) are: Na$_4$Cu$_2$S$_3$ (*I*4$_1$/a), Na$_3$Cu$_4$S$_4$ (*Pbam*), Na$_7$Cu$_{12}$S$_{10}$ (*P*2/*m*), NaCu$_4$S$_4$ (*P*-3*m*1), Na$_2$Cu$_4$S$_3$ (*C*2/*m*), NaCu$_4$Se$_4$ (*P*6$_3$/*mmc*), NaCuSe (*P*4/*nmm*), NaCuTe (*P*4/*nmm*), NaCu$_3$Te$_2$ (*R*3*m*), K$_3$Cu$_8$S$_6$ (*C*2/*m*), KCu$_4$S$_3$ (*P*4/*mmm*), KCuS (*Pnma*), KCu$_3$S$_2$ (*C*2/*m*), KCu$_2$Se$_2$ (*I*4/*mmm*), KCuSe (*P*6$_3$/*mmc*), K$_3$Cu$_8$Se$_6$ (*C*2/*m*), KCu$_4$Se$_3$ (*P*4/*mmm*), K$_2$Cu$_5$Te$_5$ (*Cmcm*), KCuTe (*P*6$_3$/*mmc*), K$_4$Cu$_8$Te$_{11}$ (*C*2/*m*), K$_3$Cu$_{11}$Te$_{16}$ (*Imma*), KCu$_3$Te$_2$ (*C*2/*m*), K$_2$Cu$_2$Te$_5$ (*Cmcm*), RbCu$_4$S$_3$ (*P*4/*mmm*), RbCu$_3$S$_2$ (*C*2/*m*), Rb$_3$Cu$_8$S$_6$ (*C*2/*m*), and $Rb_3Cu_8Se_6$ ($C2/m$). The theoretically predicted[34] ternary compounds in A-Cu-X systems are: NaCuS ($P6_3/mmc$), RbCuS ($Cmcm$), RbCuSe ($Cmcm$), and RbCuTe ($Pnma$). Most of the achiral alkali metal copper chalcogenide members are centrosymmetric compounds exhibiting hidden spin polarization effects [1]. Among them NaCuS, KCuSe, and KCuTe are predicted to have hidden Dresselhaus (D-2) effect [1], the rest are predicted to have hidden Rashba (R-2 & D-2) effect [1]. On the basis of structural symmetry analysis, we have found one compound $NaCu_3Te_2$ ($R3m$) with a polar space group permitting Rashba spin splitting.

**Appendix III. Electronic structures of the chiral compound $NaCu_5S_3$ without SOC**

The crystallographic information of the chiral alkali metal copper chalcogenide compound, $NaCu_5S_3$ ($P6_322$), is given in Table A1, along with the corresponding AC structure in space group $P6_3/mcm$. The chiral structure ($P6_322$) given in Table A1 corresponds the FEC-1 structure in the main text, which is the experimental phase. Table A2 lists the lattice constants of the potential substrate InN for applying tensile strain on $NaCu_5S_3$.

Fig. A1 shows the evaluated band structure for $NaCu_5S_3$ from DFT+$U$ ($U$ = 6 eV) without SOC for comparison with the band structure with SOC in Fig. 5(a, b) and Fig. A2 shows the partial density of states and charge densities of the FEC-1/2 and AC phases in $NaCu_5S_3$ from DFT+$U$ without SOC. The top valence bands of the three phases are mainly formed by hybridized S ($3p_x/p_y$) and Cu ($3d$) states. In the charge densities of $NaCu_5S_3$, there are four empty interstitial sites per unit cell that have the same size for the AC structure. The sizes of the four interstitial sites of the FEC-1/2 structures disproportionate, i.e. half of them increases and the other half decreases [Fig. A2b]. This small change of atomic coordination as well as charge distribution leads to the ferri-chirality. The calculated band gap of $NaCu_5S_3$ from DFT+$U$=6 eV is almost the same as the band gap from HSE06 (~0.12 eV). Fig. A3 compares the band structures obtained with the two different functionals. The band splitting of the top valence bands is weakly dependent on the $U$ parameter [Fig. A4].

**Table A1.** Wyckoff positions of $NaCu_5S_3$ with $P6_322$ and $P6_3/mcm$ symmetry.

| Space group | Site | Wyckoff position | $x$ | $y$ | $z$ |
| --- | --- | --- | --- | --- | --- |
| $P6_322$ | Na | 2b | 0 | 0 | 1/4 |
| | Cu | 6h | -0.4854 | 0.4854 | 1/4 |
| | Cu | 4f | 2/3 | 1/3 | -0.0247 |
| | S | 6g | 0.6699 | 0.6699 | 1/2 |
| $P6_3/mcm$ | Na | 2b | 0 | 0 | 1/2 |
| | Cu | 6f | 1/2 | 1/2 | 1/2 |
| | Cu | 4c | 2/3 | 1/3 | 1/4 |
| | S | 6g | 0.672 | 0.672 | 3/4 |

**Table A2.** Comparison of lattice constants of potential substrate InN and NaCu$_5$S$_3$, where the $a_{sub}$ and $a_{sub}^{opt}$ represents the primitive and minimal coincident lattice constants of specific crystal plane.

| | $a_{sub}$ (Å) | $a_{sub}^{opt}$ (Å) | Mismatch (%) $\|a_{sub}^{opt} - a_{NaCu_5S_3}\| / a_{NaCu_5S_3}$ | |
|---|---|---|---|---|
| wz-InN (001) | 3.584 | 7.168 (2×2) | +2.3 | Tensile |

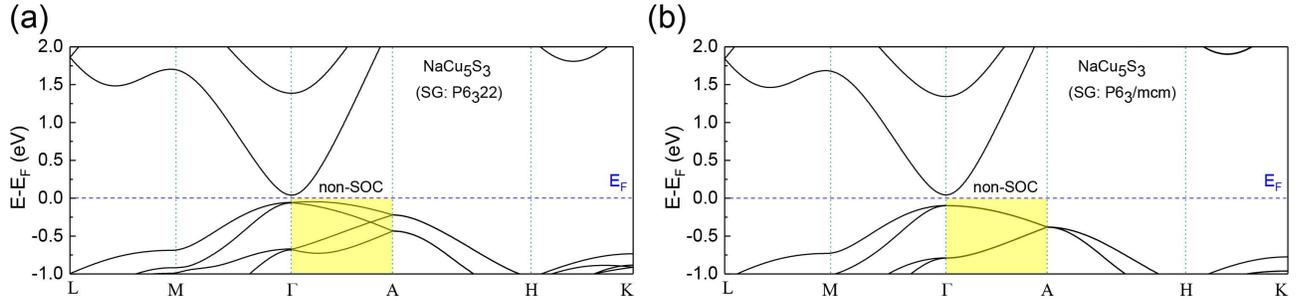

**FIG. A1.** (a) Band structure of NaCu$_5$S$_3$ with $P6_322$ symmetry. (b) Band structure of NaCu$_5$S$_3$ with $P6_3$/mcm symmetry. Here the band structures were obtained from DFT+$U$ ($U$ = 6 eV) without SOC.

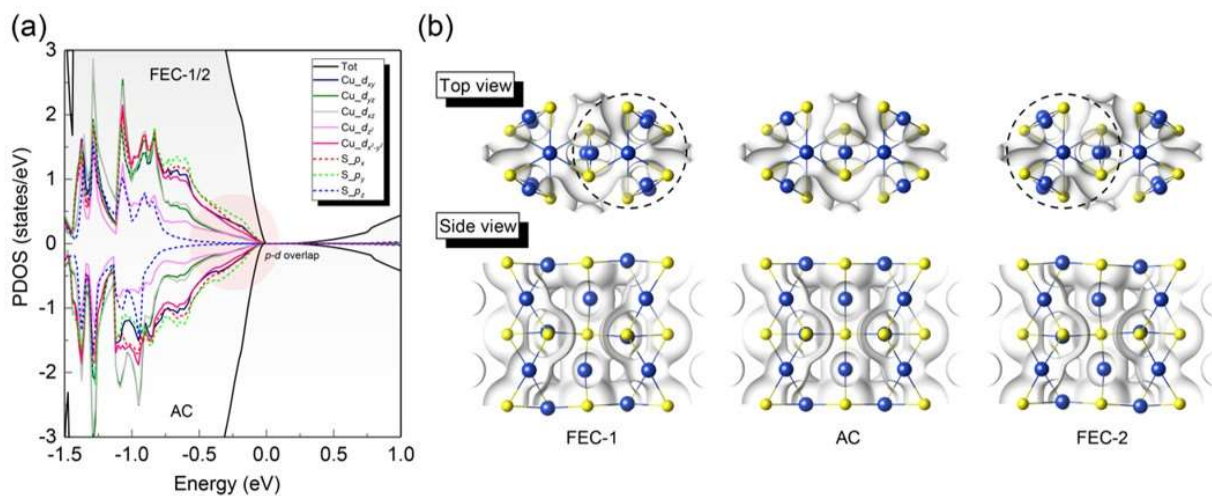

**FIG. A2.** (a) Partial density of states (PDOS) for NaCu$_5$S$_3$ with $P6_322$ (FEC-1/2) and $P6_3/mcm$ (AC) symmetries from DFT+$U$ ($U$ = 6 eV) without SOC. (b) Charge density ($\rho$=0.02 eÅ$^{-3}$) of the FEC-1/2 and AC phases.

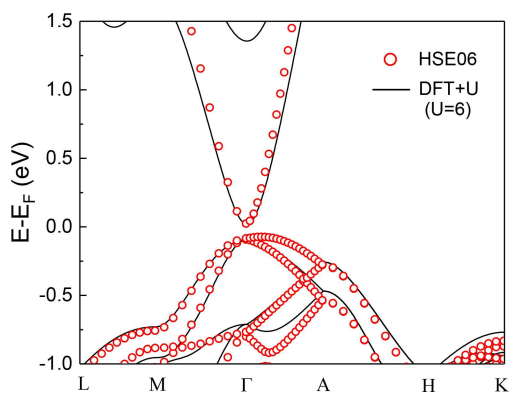

**FIG. A3.** Comparison of band structures of NaCu$_5$S$_3$ with $P6_322$ symmetry from HSE06 and DFT+$U$ ($U$ = 6 eV), without SOC.

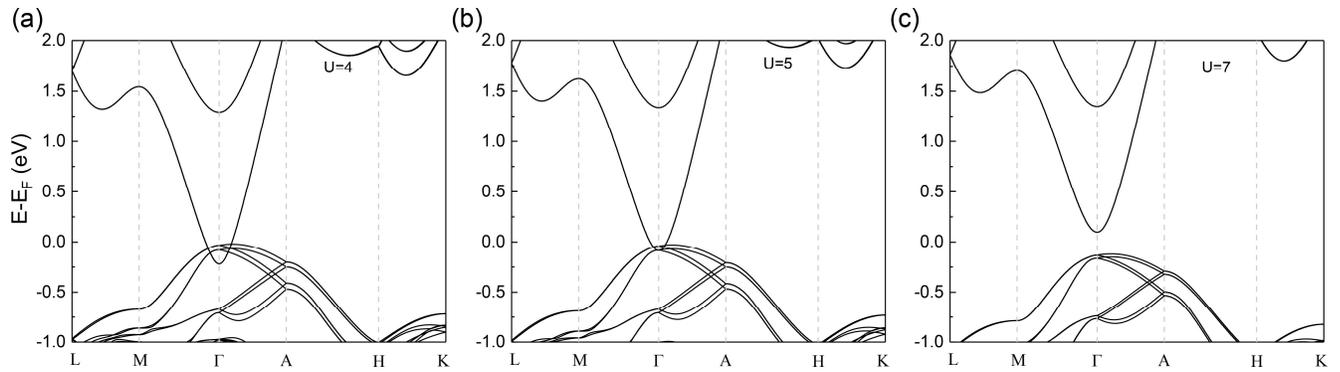

**FIG. A4.** Band structures of NaCu$_5$S$_3$ with $P6_3 22$ symmetry from DFT+$U$ without SOC with different U values: (a) $U$ = 4 eV, (b) $U$ = 5 eV, (c) $U$ = 7 eV.